\documentclass[twocolumn]{aastex701} 

\usepackage[T1]{fontenc}
\usepackage{ae,aecompl}
\usepackage{hyperref}
\usepackage{url}

\usepackage{longtable}
\usepackage{booktabs}
\usepackage{tabu}
\usepackage{graphicx}
\usepackage{multirow}
\usepackage{ulem}
\usepackage{color}
\usepackage{amsmath}

\def \s{~\rm{s}}
\def \km{~\rm{km}}

\def \K{~\rm{K}}

\def \g{~\rm{g}}
\def \G{~\rm{G}}

\def \erg{~\rm{erg}}

\def \yr{~\rm{yr}}

\def \pc{~\rm{pc}}
\def \kpc{~\rm{kpc}}

\usepackage{xcolor}
\definecolor{redak}{rgb}{0.9,0.15,0.05}
\shorttitle{Attributing the SNR RCW 89 to the JJEM}
\shortauthors{Soker}

\graphicspath{{./}{figures/}}

\begin{document}

\title{Attributing the supernova remnant RCW 89 to the jittering jets explosion mechanism}

\author[0000-0003-0375-8987]{Noam Soker} 
\affiliation{Department of Physics, Technion Israel Institute of Technology, Haifa, 3200003, Israel; soker@physics.technion.ac.il}
\email{soker@physics.technion.ac.il}

%\date{\today}

\begin{abstract}
I examine recent radio observations of the supernova remnant (SNR) RCW 89 and identify a point-symmetric morphology composed of two main symmetry axes. I attribute this morphology to a jet-driven explosion in the framework of the jittering jets explosion mechanism (JJEM). To reach this conclusion, I argue that the MSH 15-52 nebula associated with the pulsar PSR B1509-58, the X-ray hand-like-shaped nebula, and RCW 89 are two separate core-collapse supernova (CCSN) remnants that interact with each other. Namely, the nebula SNR G320.4-1.2 contains two CCSN remnants. In essence, I utilize the recent successes of the JJEM to account for the morphologies of point-symmetric CCSN remnants, thereby explaining the morphology of RCW 89 and identifying it as a separate CCSN remnant. I suggest a process by which somewhat more energetic pairs of jets in the JJEM have a positive feedback on the accreted gas onto the newly born neutron star, thereby prolonging the life of the jets and explaining the occurrence of two or three energetic pairs of jets in some CCSNe. This study adds RCW 89 to the growing list of point-symmetric CCSN remnants. The JJEM naturally explains these morphologies as shaped by misaligned pairs of jets that exploded these CCSNe.    
\end{abstract}


% ==================================
\section{Introduction} 
\label{sec:intro}
% ==================================

There is a fierce debate between the supporters of the two actively studied alternative theoretical explosion mechanisms of core-collapse supernova (CCSN) (as evident, for example, in two supernova meetings in 2025): the jittering-jets explosion mechanism (JJEM; \citealt{Soker2025Padova}\footnote{\url{https://www.memsait.it/videomemorie/volume-2-2025/VIDEOMEM_2_2025.47.mp4}}), and the neutrino-driven explosion mechanism (e.g., \citealt{Janka2025Padova}\footnote{\url{https://www.memsait.it/videomemorie/volume-2-2025/VIDEOMEM_2_2025.46.mp4} and \url{https://www.youtube.com/watch?v=nRfDPPSmnzI&list=PLfyHSCP2-a1VyU2S4gSOI4Wta1uC1MX9H&index=1&t=100s}}). 
Only one research group studies the JJEM (and, therefore, one of my goals is to convince more in the community to study the JJEM). However, on the other hand, the groups studying the neutrino-driven mechanism do not agree on even qualitative results (see the meeting recordings above and \citealt{Soker2024UnivReview} for the current status of the debate). 

Studies of the delayed neutrino explosion mechanism focused on simulating the collapse of the core, the revival of the stalled shock by neutrino heating, comparing to some observations, and in finding the conditions for a star to explode (e.g., \citealt{Bambaetal2025CasA, Bocciolietal2025, EggenbergerAndersenetal2025, Huangetal2025, Imashevaetal2025, Laplaceetal2025, Maltsevetal2025, Maunderetal2025, Morietal2025, Mulleretal2025, Nakamuraetal2025, SykesMuller2025, Janka2025, Orlandoetal20251987A, ParadisoCoughlin2025, Tsunaetal2025, Vinketal2025, WangBurrows2025, Willcoxetal2025}). The magnetorotational explosion mechanism involves the launching of jets along a fixed axis because of rare cases with rapidly rotating pre-collapse cores (e.g., \citealt{Shibagakietal2024, ZhaMullerPowell2024, Shibataetal2025}). The magnetorotational explosion mechanism attributes most CCSNe to the neutrino-driven mechanism, and, therefore, I group them together.

The fundamental process of the JJEM is the launching of $N_{\rm 2j} \simeq 5-30$ pairs of jittering jets by intermittent accretion disks (or belts) around the very young neutron star (NS; \citealt{Soker2025Learning} for a recent review of the JJEM parameters and differences from the neutrino-driven mechanism). \cite{braudoetal2025} conduct three-dimensional hydrodynamical simulations of the shaping of the ejecta by jittering jets, and \cite{SokerAkashi2025} of the shaping by a pair of jets at a late phase of the explosion process.  
The two jets in a pair might substantially differ in their opening angle and power (e.g., \citealt{Bearetal2025Puppis}), and might be at an angle smaller than $180^\circ$, namely, not exactly opposite (e.g., \citealt{Shishkinetal2025S147}). I return to some properties of the jittering jets in Section \ref{sec:PowrfulPair}.

The stochastic angular momentum variations of the mass that the newly-born NS accretes result from the stochastic convection motion in the convective zones in the pre-collapse core that turbulence and instabilities above the NS amplify (e.g., \citealt{GilkisSoker2014, GilkisSoker2016, ShishkinSoker2021, ShishkinSoker2023, WangShishkinSoker2024, WangShishkinSoker2025}; for the studies of the turbulence and instability irrespective of the JJEM, see, e.g., \citealt{Abdikamalovetal2016, KazeroniAbdikamalov2020, Buelletetal2023}). One achievement of the simulations of the neutrino-driven mechanism is demonstrating the process of neutrino heating of material below the stalled shock, the gain region. Once the jets start to explode the core, neutrino heating boosts the explosion by adding energy \citep{Soker2022nu}; however, the jittering jets supply most of the energy for the explosion. 
I also point out that many models of observed superluminous CCSNe add a magnetar to supply extra energy to the explosion. In most of these cases, jets supply more energy than the magnetar \citep{Soker2022magnetar}. Moreover, jets, i.e., the JJEM, explode the star rather than the neutrino-driven mechanism that cannot supply the required explosion energy of $E_{\rm exp} \gtrsim 3 \times 10^{51} \erg$ of many superluminous CCSNe (e.g., \citealt{SokerGilkis2017, Kumar2025}).

The rich variety of processes that occur during the formation and evolution of supernova remnants (SNRs) can reveal the explosion mechanism and pre-explosion mass loss. These processes include emission of radiation including cosmic rays (e.g., \citealt{Yamazakietal2014RAA, Zhangetal2016RAA, Lietal2020RAA, Luoetal2024RAA}), dust in the ejecta (e.g., \citealt{Shahbandehetal2023, Shahbandehetal2025, LuXiWeietal2025RAA}), the role of the NS remnant in the center (e.g., \citealt{HorvathAllen2011RAA, Wuetal2021RAA}), magnetohydrodynamics (e.g., \citealt{Wuetal2019RAA, Leietal2024RAA}), and ejecta-ambient gas interaction (e.g., \citealt{Yanetal2020RAA, Luetal2021RAA}) including jets (e.g., \citealt{YuFang2018RAA}). The most relevant to the present goals is the morphology of the ejecta, e.g., the paper cited below on point symmetrical CCSN remnants (CCSNRs; see also, e.g., \citealt{Renetal2018RAA} on non-point-symmetric morphologies). This study examines the CCSNR morphology of a specific CCSNR, RCW 89. 

The identification of point-symmetric morphologies in approximately fifteen CCSN remnants (CCSNRs), mostly in 2024-2025, led to a breakthrough in the exploration of the JJEM. 
Most recent point-symmetric morphology identifications, with references to earlier identifications, are: Puppis A \citep{Bearetal2025Puppis}, SNR G0.9+0.1 \citep{Soker2025G0901}, S147 \citep{Shishkinetal2025S147}, and N132D \citep{Soker2025N132D}.  
Point-symmetric morphologies are those that have two or more pairs of opposite structural features that are not along the same axis; structural features include clumps, filaments, bubbles, lobes, ears, and rings. The JJEM predicts point-symmetric morphologies in many, but not all, CCSNe (see detailed discussion by \citealt{SokerShishkin2025Vela}); the existence of many point-symmetric CCSNRs is a severe challenge to the neutrino-driven mechanism.     
Any new CCSNR with an identified point-symmetric morphology strengthens the JJEM. With the breakthrough in establishing the JJEM as the primary explosion mechanism of CCSNe, the task is no longer only to identify a point-symmetric morphology to strengthen the JJEM, but also to use the successes of the JJEM in explaining properties of CCSNe and CCSNRs, and to search each CCSNR for the signatures of jittering jets, mainly point-symmetric morphologies. 
In this study, I use the JJEM to identify a point-symmetric morphology in the CCSNR RCW 89, and use this to suggest that the pulsar PSR B1509-58 and its hand-like-shaped `cosmic hand'  nebula are a separate CCSN from RCW 89. However, the two are likely interacting (Section \ref{sec:PointSymmetry}).
I summarize in Section \ref{sec:Summary}

% =========================
\section{Point-symmetry in SNR RCW 89}
\label{sec:PointSymmetry}
% =========================

% =========================
\subsection{Some early studies of G320.4-01.2}
\label{subsec:Early}
% =========================

The diffuse object G320.4-01.2 (e.g., \citealt{ShaverGoss1970, Milne1972, MilnrDickel1975}) includes the CCSNR RCW 89, as appears in H$\alpha$ (e.g., \citealt{vandenBergh1978}) and radio, and the pulsar wind nebulae of the pulsar PSR B1509-58 (MSH 15-52;  e.g.,  \citealt{BrazierBecker1997, Gaensleretal1999}).\footnote{\url{https://chandra.si.edu/photo/2025/msh1552/}} 
Following \cite{Gaensleretal1999}, most studies consider PSR B1509-58 and RCW 89 as one CCSNR (e.g., \citealt{Chevalier2005} who worked out the energetics of the system). 
In this study, I focus solely on the morphology of RCW 89, located north of the pulsar, and therefore do not refer to the pulsar or its properties (many papers study the pulsar and its surroundings, e.g., \citealt{Manchesteretal1982, SewardHarnden1982, Weisskopfetal1983, Aharonianetal2005, Yatsuetal2009, Abdoetal2010, Romanietal2023}). I accept the conclusions of \cite{Gaensleretal1999} that RCW 89 and the nebula of PSR B1509-58 are at the same distance and have a similar age, and interact with each other. However, based on the morphology of RCW 89 in the framework of the JJEM, I raise the possibility that these are two separate CCSNRs. 

% =========================
\subsection{Identifying RCW as a point-symmetric CCSNR}
\label{subsec:RCW89}
% =========================

\begin{figure*}[]
	\begin{center}
%	\hspace*{-2cm} 
	% This cut edges: [trim=left bottom right top, clip]{file}
%	\hspace{1cm}
\includegraphics[trim=0.0cm 18.9cm 0.0cm 0.0cm ,clip, scale=0.85]{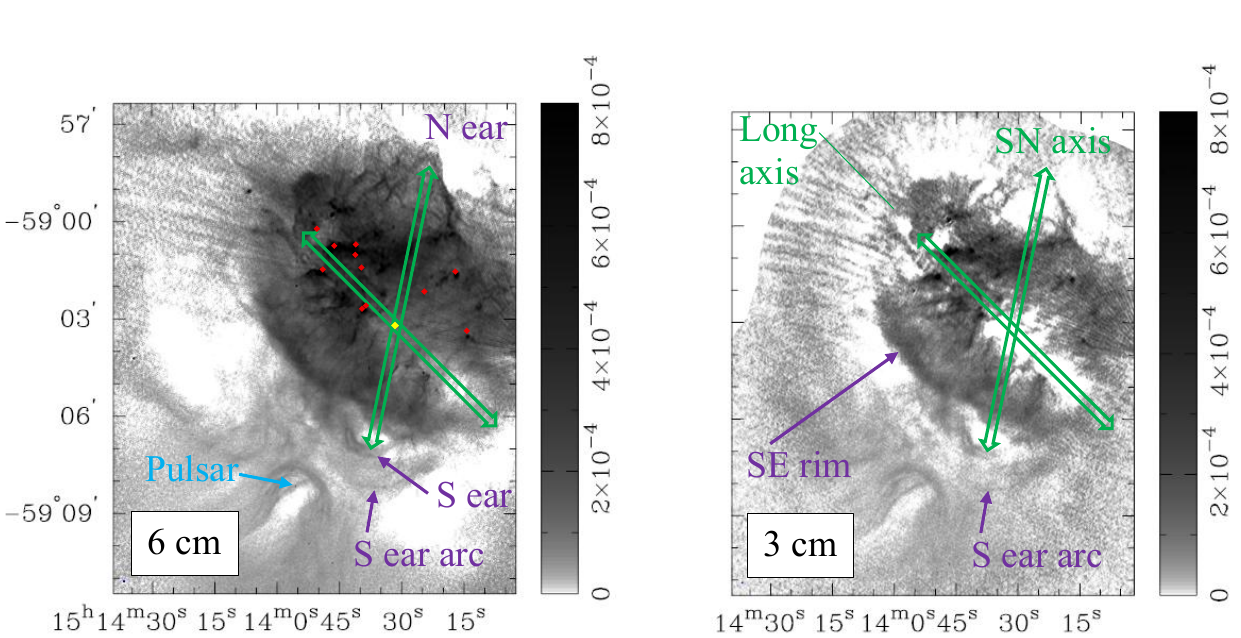} 
\caption{
Images adapted from \cite{Zhangetal2025} showing the total intensity radio maps of MSH 15-52 and RCW 89 at 6 cm (left) and 3 cm (right). The radio emission extends to the northeast (Figure \ref{Fig:RCW89FigureRadioExtend}). The grayscale bars have units of Jy/beam. Axes are RA and DEC (J2000). I identify and mark several morphological features, as well as the two symmetry axes that I attribute to shaping by two pairs of energetic jets that are part of the pairs of jets that exploded the star in the framework of the JJEM. The red dots in the left panel mark the location of the X-ray knots for which \cite{Borkowskietal2020} measure the proper motion.  
}
%\vskip+0.5cm
\label{Fig:RCW89FigureRadioCen}
\end{center}
\end{figure*}

\begin{figure}[]
	\begin{center}
%	\hspace*{-2cm} 
	% This cut edges: [trim=left bottom right top, clip]{file}
%	\hspace{1cm}
\includegraphics[trim=1.9cm 17.9cm 0.0cm 0.0cm ,clip, scale=0.81]{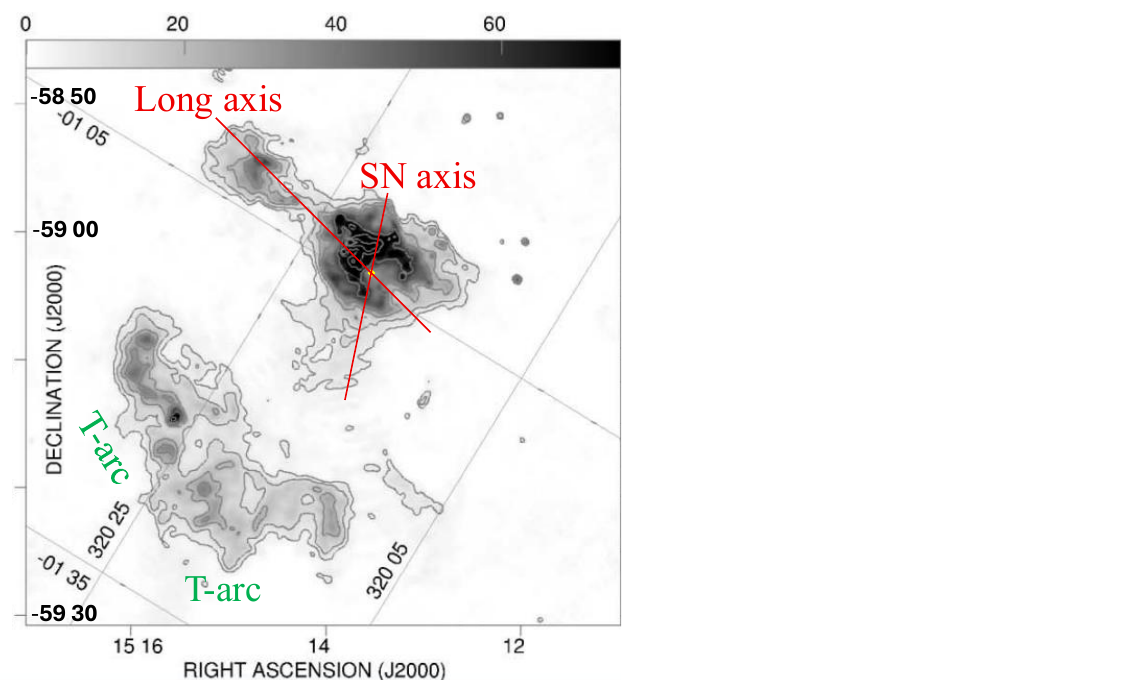} 
\caption{A radio continuum image pf G320.4-1.2 at 1.4 GHz adapted from \cite{Dubneretal2002}. The gray scale is mJy/beam. Contours levels are 5, 10, 20, 30, 60, 90, 120, and 150 mJy/beam. The diagonal lines are Galactic coordinate lines. 
I added two lines along the same two symmetry axes that I identify in Figure \ref{Fig:RCW89FigureRadioCen}. The long symmetry axis is at $14^\circ$ to the Galactic plane.  
}
%\vskip+0.5cm
\label{Fig:RCW89FigureRadioExtend}
\end{center}
\end{figure}

Although earlier studies of the morphology of G320.4-1.2 exist, such as \cite{Caswelletal1981} in radio and \cite{Gaensleretal2002} and \cite{Dubneretal2002}, which compared radio and X-ray morphologies, the new publication by \cite{Zhangetal2025} motivates the present study. In Figure \ref{Fig:RCW89FigureRadioCen}, I present two images from \cite{Zhangetal2025} at two radio wavelengths. Based on earlier studies (e.g., \citealt{Soker2024NA1987A, Soker2024PNSN, Bearetal2025Puppis, Soker2025G0901, SokerShishkin2025W49B}) that compared the morphologies of some jet-shaped planetary nebulae with some other CCSNRs, I identify two symmetry axes. The long symmetry axis is aligned along the long dimension of RCW 89, extending from a radio-faint zone in the southwest to a radio-faint zone in the northeast. The other axis, the SN axis, connects the south ear (S ear) to the north ear (N ear). I also identify an arc outside the south ear, the S ear arc.
In Figure \ref{Fig:RCW89FigureRadioExtend}, I mark the two symmetry axes on a more extended radio image that I adapted from \cite{Dubneretal2002}. Although I identify the long axis by the radio-faint zone in the main RCW 89 nebula, it crosses through the peak emission to the northeast of the main nebula.  I note that the long axis of RCW 89 is at $14^\circ$ to the galactic plane. I attribute the shaping of the two symmetry axes to two energetic pairs of jets.  
% FFFFFFFFFFFFFFFFFFFFFFFFFFFFFFFFFFFFFFFFFF
% FFFFFFFFFFFFFFFFFFFFFFFFFFFFFFFFFFFFFFFFFFF
% FFFFFFFFFFFFFFFFFFFFFFFFFFFFFFFFFFFFFFFFFFF
% FFFFFFFFFFFFFFFFFFFFFFFFFFFFFFFFFFFFFFFFFff

\begin{figure}[]
	\begin{center}
%	\hspace*{-2cm} 
	% This cut edges: [trim=left bottom right top, clip]{file}
%	\hspace{1cm}
\includegraphics[trim=0.0cm 0.0cm 0.0cm 0.0cm ,clip, scale=0.53]{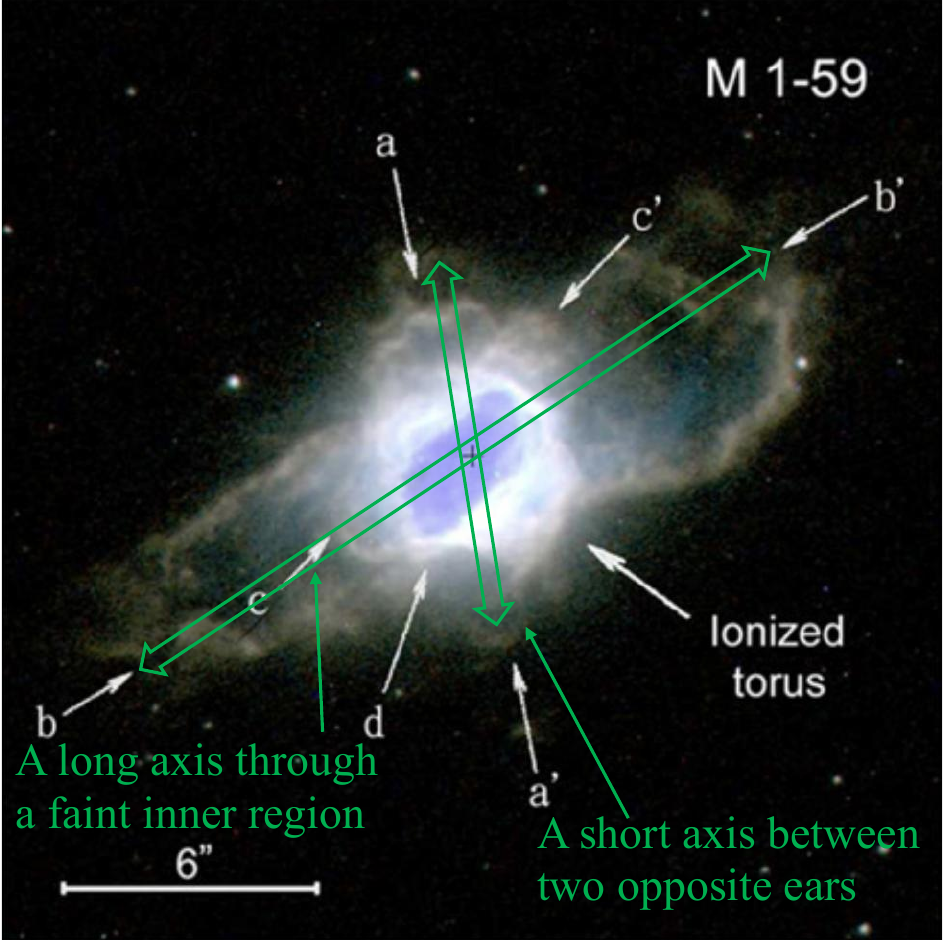} 
	\vspace*{-7.5cm} 
\caption{{ An image of the point-symmetric (multipolar) planetary nebula M~1-59 adapted from \cite{Hsiaetal2014}, who identified the pairs a-a', b-b', and c-c', and attributed the shaping to pairs of jets (bipolar outflows). I added the two green double-lined, two-sided arrows to mark two axes on the morphology of M~1-59.  }
}
\label{Fig:M159}
\end{center}
\end{figure}

{ Although the identification of two jet-shaped axes contains a speculative component, the identification is based on many jet-shaped planetary nebulae that have morphologies with similar morphological features, and which researchers of planetary nebulae attribute the shaping to pairs of jets (bipolar outflows). I therefore consider the identification of the two axes as robust. In Figure \ref{Fig:M159} I present one point-symmetric planetary nebula (termed multipolar planetary nebula), M 1-59 from \cite{Hsiaetal2014}. They identified the point-symmetric morphology of M~1-59 by pairs of structural features that they mark by white letters on the figure: a-a', b-b', and c-c'. I identify in Figure \ref{Fig:M159} two axes (out of three or even four) that resemble those of RCW 89. 
The long axis coincides with the long dimension of M~1-59 and crosses the elongated faint zone in the center. This resembles the long axis of RCW 89, which also passes through a faint region (in radio) and coincides with the long dimension of the SNR. The short axis of the planetary nebula M~1-59 connects two ears that \cite{Hsiaetal2014} mark. This resembles the two ears and the SN axis they define in RCW 89. The point of this comparison between RCW 89 and the planetary nebula M~1-59 is that the same tools that researchers apply to identify jet-shaped point-symmetric structures in planetary nebulae (e.g., \citealt{Lopezetal2000, Sahai2000, Sahaietal20112011, Sabinetal2017, Bandyopadhyayetal2023, Wenetal2024}) are used here to identify the two axes of RCW~89. }
% FFFFFFFFFFFFFFFFFFFFFFFFFFFFFFFFFFFFFFFFFFF
% FFFFFFFFFFFFFFFFFFFFFFFFFFFFFFFFFFFFFFFFFff

\begin{figure}[]
	\begin{center}
%	\hspace*{-2cm} 
	% This cut edges: [trim=left bottom right top, clip]{file}
%	\hspace{1cm}
\includegraphics[trim=0.0cm 0.5cm 9.0cm 10.7cm ,clip, scale=0.75]{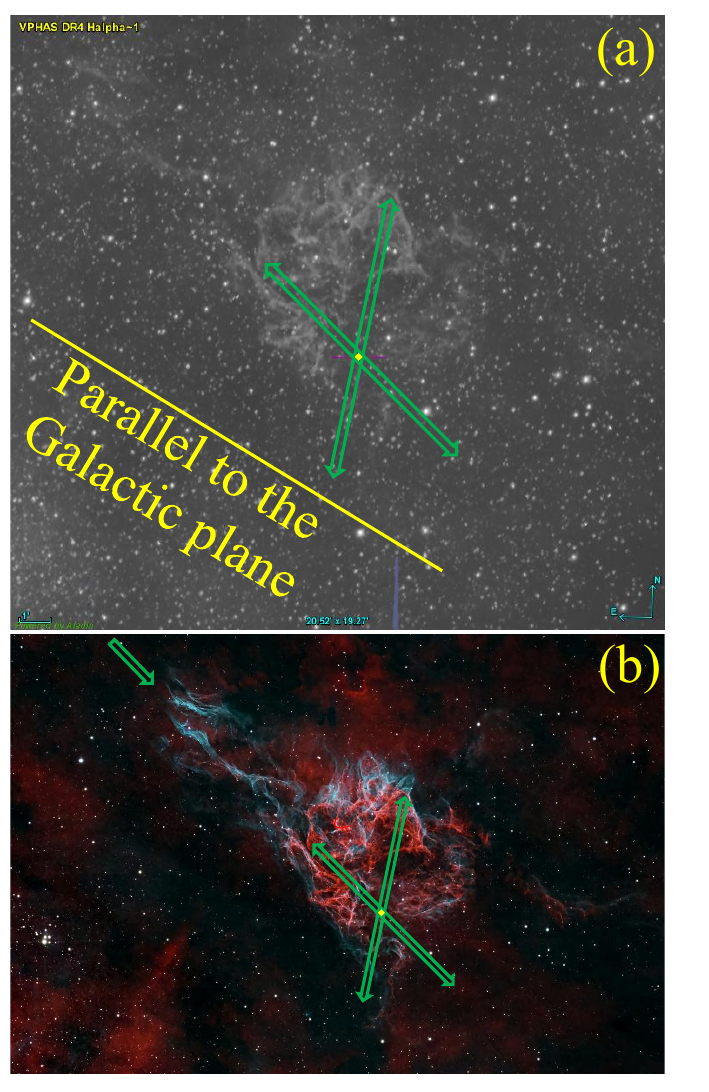} 
\caption{Two images of RCW 89 with the two double-sided arrows in the same place and scale as in Figure \ref{Fig:RCW89FigureRadioCen}. The images are not on the same scale. (a) An H$\alpha$ image of VPHAS Data created with Aladin   (\citealt{Aladin1, Aladin2, aladin3}). 
(b) An image processed and produced by Manuel C. Peitsch (Roboscopes at e-EyE Entre Encinas y Estrellas Observatory, Camino de los Molinos, 06340 Fregenal de la Sierra, Spain:  \url{https://manuel-astro.ch/}) based on original observations by Martin R. Pugh (Observatorio El Sauce, Rio Hurtado, Coquimbo, Chile). Red represents H$\alpha$ emission and blue-green represents [O \textsc{iii}] emission. The arrow on the upper left is along the line of the double-sided arrow of the long axis in the center. 
}
%\vskip+0.5cm
\label{Fig:RCW89Optical}
\end{center}
\end{figure}

I identify the RCW 89 as a full CCSNR, rather than part of a CCSN ejected by the progenitor of the pulsar PSR B1509-58 to the south, for the following arguments. 
\begin{enumerate}
\item The SE rim, the S ear arc, and the S ear (as I mark in Figure \ref{Fig:RCW89FigureRadioCen}) have morphologies that suggest outflows to the south and southeast from the center of RCW 89 (the intersection of the two green-double-sided arrows), rather than a flow from the pulsar to their directions. 
\item The two symmetry axes that I identify form a two-axis point-symmetric morphology similar in some respects to other CCSNRs with identified point-symmetric morphologies, e.g, Cassiopeia A with its pronounced jet to the east and two opposite arcs \citep{BearSoker2025}, CCSNR W44 with its two main symmetry axes \citep{Soker2024W44}, The Cygnus Loop with its three main symmetry axes, one of them identified by two opposite arcs \citep{ShishkinKayeSoker2024}, and Puppis A with its unequal sides and inclined pairs of ears \citep{Bearetal2025Puppis}. 
\item The long axis that the inner bright radio structure defines in Figure \ref{Fig:RCW89FigureRadioCen}, continuous into the bright radio peak in the northeast radio extension (Figure \ref{Fig:RCW89FigureRadioExtend}), which coincides with the elongated H$\alpha$ structure seen in panel (a) and in red in panel (b) of Figure \ref{Fig:RCW89Optical}, and the [O~\textsc{iii}] elongated structure seen in blue in panel (b) of Figure \ref{Fig:RCW89Optical}.  I attribute the extended northeast structure to a jet that was one of two jets of a pair that shaped the long axis, and this pair was one of several pairs of jets that exploded the progenitor of RCW 89. 
\end{enumerate} 
% FFFFFFFFFFFFFFFFFFFFFFFFFFFFFFFFFFFFFFFFFF
% FFFFFFFFFFFFFFFFFFFFFFFFFFFFFFFFFFFFFFFFFFF

Figures \ref{Fig:RCW89FigureRadioCen} - \ref{Fig:RCW89Optical} and the arguments above comprise my claim that RCW 89 is a separate CCSNR than the nebula of pulsar PSR B1509-58. I turn to discuss their possible interaction. 

% =========================
\subsection{Interaction with the pulsar wind nebula}
\label{subsec:PWN}
% =========================

\begin{figure}[]
	\begin{center}
%	\hspace*{-2cm} 
	% This cut edges: [trim=left bottom right top, clip]{file}
%	\hspace{1cm}
\includegraphics[trim=1.1cm 16.4cm 0.0cm 0.0cm ,clip, scale=0.64]{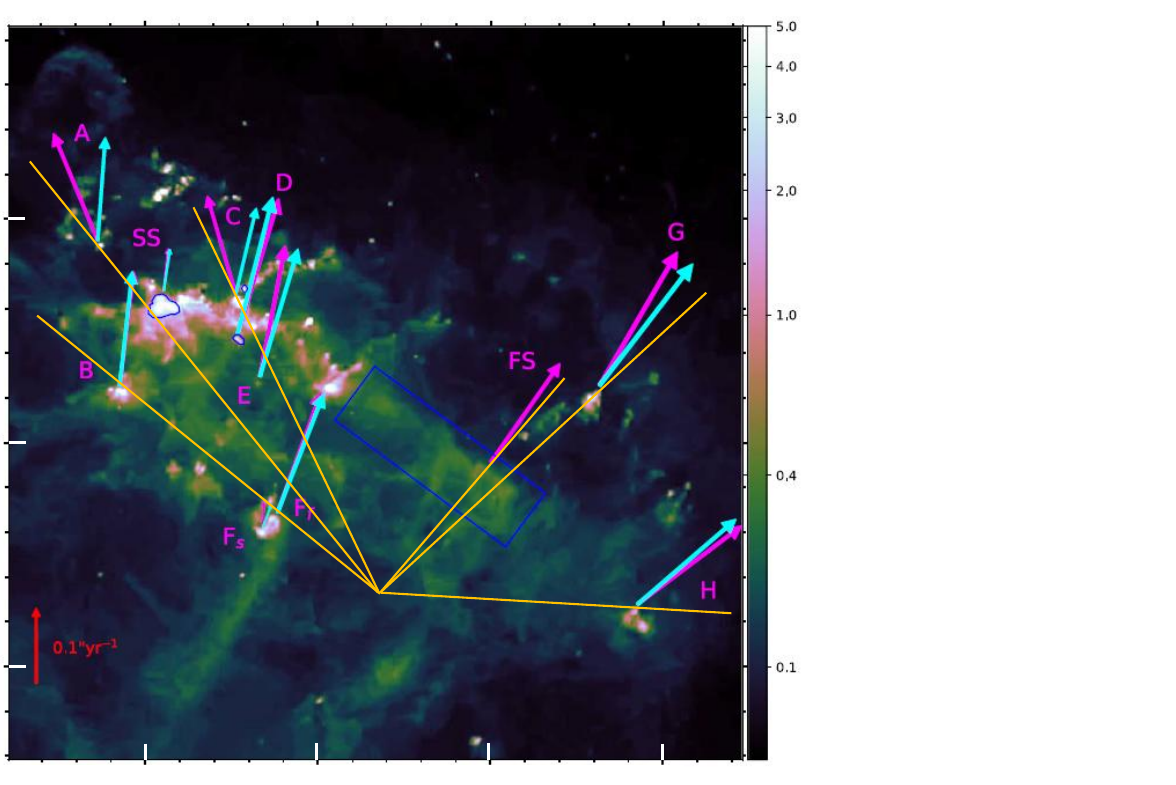} 
\caption{An image adapted from \cite{Borkowskietal2020} showing the proper motion of X-ray bright knots (red dots mark these knots on the radio map in Figure \ref{Fig:RCW89FigureRadioCen}). Magenta arrows show the proper motion, and cyan arrows are the velocity components with respect to the pulsar. (The regions encircled in blue were used by them to extract spectra.) 
Axes are RA with large ticks of $15^{\rm h}13^{\rm m}48^{\rm s}$ to $15^{\rm h}13^{\rm m}12^{\rm s}$ (left to right), and DEC with large ticks of $-59^{\circ}04^{\prime}$, $-59^{\circ}02^{\prime}$, and $-59^{\circ}00^{\prime}$ (bottom up). Red arrow on the lower left is the scale for $0.1^{\prime \prime} \yr^{-1}$. I added the six orange lines from the center of RCW 89 that I identified by the intersection of the two two-sided arrows in Figure \ref{Fig:RCW89FigureRadioCen}.
}
%\vskip+0.5cm
\label{Fig:RCW89Velocity}
\end{center}
\end{figure}

\cite{Borkowskietal2020} found the proper motion of the X-ray knots to be in the range of $0.03-0.2~^{\prime \prime}\yr^{-1}$, with nine out of the eleven knots with $>0.1^{\prime \prime}\yr^{-1}$, and a median value of $0.16^{\prime \prime}\yr^{-1}$. On the other hand, \cite{vandenBerghKsamper1984} find 13 optical knots to have slower expansion, with a proper motion component from the pulsar in the range of $-0.74^{\prime \prime}\yr^{-1}$ (i.e., towards the pulsar) to $+0.79^{\prime \prime}\yr^{-1}$. The two optical filaments with the fastest proper motion \textit{towards the pulsar} are the closest to it, and generally moving away from the center of RCW 89. The proper motions of these two filaments are counter to an explosion from the pulsar, but compatible with an explosion at the center of RCW 89 (the intersection of the two double-sided arrows in figure \ref{Fig:RCW89FigureRadioCen}). 
\cite{Borkowskietal2020} also found the directions of motion of the X-ray knots they studied to expand away from PSR B1509-58 generally. Figure \ref{Fig:RCW89Velocity} that I adapted from \cite{Borkowskietal2020}, presents the proper motion of X-ray knots in magenta, and their component from the Pulsar in cyan.  I draw the six orange lines from the center of RCW 89; the center that I identified in Figure \ref{Fig:RCW89FigureRadioCen}, i.e., the intersection of the two double-sided arrows there. The proper motion of knots A and C is at a high angle from the line to the pulsar, and much closer to the line to the center of RCW 89 (the respective orange lines). The proper motions of some knots point directly at the pulsar (e.g., B and FS). 
The differences in the proper motions of the X-ray knots and the optical ones further suggest that RCW 89 is a separate CCSNR than the nebula associated with the pulsar SR B1509-58. 
% FFFFFFFFFFFFFFFFFFFFFFFFFFFFFFFFFFFFFFFFFF
% FFFFFFFFFFFFFFFFFFFFFFFFFFFFFFFFFFFFFFFFFff

I find, nevertheless, indications that RCW 89 and the pulsar nebula interact with each other. In the optical (Figure \ref{Fig:RCW89Optical}), the RCW 89 CCSNR is much smaller in the side facing the pulsar than on the other side. I suggest that the CCSNR of PSR B1509-58 compressed the optically-emitting parts of RCW 89 towards the northwest. The same interaction forms and influences the X-ray knots and their proper motion that \cite{Borkowskietal2020} studied (Figure \ref{Fig:RCW89Velocity}). Their velocities result from the explosion of RCW 89 and the expanding remnant of the PSR B1509-58 CCSNR, explaining some knots with velocities away from PSR B1509-58 and others from a direction between PSR B1509-58 and the center of RCW 89. This interaction had suffered vigorous Rayleigh-Taylor instabilities, explaining the hand-like-shaped `cosmic hand' nebula that appears in X-rays. 

{ I emphasize the implication of the velocities of the X-ray knots to my suggestion of an interaction between the two CCSNe. The left panel of Figure \ref{Fig:RCW89FigureRadioCen} presents the X-ray knots on the radio image of RCW 89, while Figure \ref{Fig:RCW89Velocity} presents their velocities. The velocities do not all point back at the pulsar, nor at the center of RCW 89. Some do point at the pulsar. I suggest that the knots originate as denser parcels of gas in the explosion of the CCSNR of PSR B1509-58, but were further shaped as they interacted with the ejecta from RCW 89. This interaction also changed the velocity of some parcels of gas. These knots are not inside the main CCSN shell of RCW 89, but rather closer to us or farther away from us, i.e., on the periphery of the RCW 89 main shell. Their velocity implies that a significant interaction between the two ejecta occurred.  }

More specifically, I tentatively propose the following speculative evolutionary scenario. 
\newline (1) For tens of thousands of years, the progenitor of RCW 89 blew an intense wind that filled the space around the progenitors of RCW 89 and the progenitor of the pulsar PSR B1509-58 to the south. 
\newline (2) The progenitor of PSR B1509-58 exploded about 2000 thousand years ago, as the age of the pulsar (e.g., \citealt{Gaensleretal1999} for the age). The interaction of this CCSNR with the surrounding wind of the RCW 89 progenitor was unstable, leading to the formation of Rayleigh-Taylor instability fingers in the region between them; these instability fingers will form the fingers of the `cosmic hand.' 
\newline (3) Several hundred years after the first CCSN, the progenitor of RCW 89 exploded as an energetic CCSNR with fast ejecta. With an ejecta velocity of $\simeq 10^4 \km \s^{-1}$ in the ejecta front, in the $\simeq 1000 \yr$ since explosion, the ejecta would reach a distance of $\simeq 10 \pc$ from the center of RCW 89. At a distance of $D \simeq 5 \kpc$ (e.g., \citealt{Gaensleretal1999}) this amounts to $\simeq 7^{\prime}$ on the plane of the sky, somewhat larger than the radius of the main RCW 89 shell as the radio emission reveals (Figure \ref{Fig:RCW89FigureRadioCen}), and the distance of RCW 89 center to the pulsar PSR B1509-58. The interaction of the ejecta of RCW 89 with the previous CCSNR of the pulsar formed the bright X-ray knots and compressed the optical nebula. Instabilities intensified, forming the fingers observed in the X-ray, and allowed some of the ejecta from RCW 89 to penetrate the pulsar nebula and flow around it, creating the tail that extends away from RCW 89 and around the pulsar. The extended region to the northeast of the main RCW 89 shell is a result of a jet that was faster than the main ejecta. 
\newline The pulsar wind nebula of PSR B1509-58 fills the cavities and forms the hard X-ray emission of the `cosmic hand.' 

The simulations of this scenario are a subject for a future study.

{{ I refer to the bright emission zone to the southeast of the pulsar, the T-arc zone that I marked in Figure \ref{Fig:RCW89FigureRadioExtend}; \cite{Gaensleretal1999} marked by `T' the brightest zone in the T-arc (the two peaks on the northeast of the arc). The T-arc is twice as far from the pulsar PSR B1509-58 as RCW 89 is, and its structure is not symmetric to RCW 89 with respect to the pulsar. I, therefore, do not consider the T-arc and RCW 89 to be two sides of one CCSNR formed by pulsar PSR B1509-58. In the scenario I propose here, the ejecta of the CCSNR of PSR B1509-58 have interacted with the ejecta of RCW 89. On the other side, the ejecta of pulsar PSR B1509-58 expanded more freely and could reach larger distances. The T-arc is a result of the interaction of the PSR B1509-58 ejecta with a circumstellar material or an interstellar cloud.     }}

Overall, I argue that, contrary to earlier claims (Section \ref{subsec:Early}), observations support the identification of RCW 89 as a separate point-symmetric CCSNR (Section \ref{subsec:RCW89}). In this subsection, I argued that RCW 89 interacts with the CCSNR that formed the PSR B1509-58. 

% =========================
\section{Launching a powerful pair of jets}
\label{sec:PowrfulPair}
% =========================

There are point-symmetric CCSNRs with several pairs of structural features much smaller than the main CCSNR shell, like the several pairs of clumps and filaments in Cassiopeia A \citep{BearSoker2025} and the Vela SNR \citep{SokerShishkin2025Vela}. Some CCSNR have two, e.g., W44 \citep{Soker2024W44}, S147 \citep{Shishkinetal2025S147}, and N132D \citep{Soker2025N132D},  or three, e.g., N63A \citep{Soker2024CounterJet} and The Cygnus Loop \citep{ShishkinKayeSoker2024}, prominent symmetry axes connecting two opposite large structures. Some other CCSNRs have less well-defined point-symmetric structures or more complicated ones. 
I identified two prominent axes that comprise the point-symmetric morphology of RCW 89. I propose a process that yields a powerful pair of jets, maintaining a constant axis for a relatively long time, i.e., $>0.1 \s$, during the explosion process.

\cite{PapishSoker2014Plan} proposed a mechanism, supported by their three-dimensional hydrodynamical simulations of jittering jets, in which two pairs of jets channel further accretion onto the newly born NS, such that consecutive bipolar jets tend to be in the same plane as the first two bipolar jet episodes. Each jet inflates a lobe that ejects core material along the lobe, i.e., from a conical zone with a half-opening angle of $\simeq 10^\circ - 60^\circ$. Therefore, the first two bipolar jet episodes eject mass mainly from the plane defined by the two bipolar axes. Accretion then proceeds from the two opposite directions more or less normal to that plane. The angular momentum of this accretion flow is perpendicular to its velocity towards the NS. Since the velocity of this accreted material is perpendicular to the plane of the first two pairs of jets, its angular momentum is in that plane. Therefore, the axis of the jets that the accretion disk of this accretion flow launches is also in the same plane as the first two pairs of jets. \cite{PapishSoker2014Plan} and \cite{BearSoker2025} suggested that this planar-jittering mechanism explains the concentration of the ejecta of Cassiopeia A in and near a plane. 

Consider a pair of jets that start more powerful than typical jets because of a large angular momentum perturbation. Further consider that the jets inflate two large, opposite bubbles that eject core material throughout almost the entire sphere, except for the thin plane between the two jet-inflated bubbles. Namely, the half-opening angle of each bubble is $>80^\circ$. The accretion flow onto the NS continues only through a thin plane, namely, the velocity of the accreted material is perpendicular to the axis of the original jets. 
This velocity direction implies that the angular momentum of this material is perpendicular to the accretion plane, which is the same direction as the axis of the original pair of jets. The accreted material can suppress the accretion disk or add to it, but it does not change its direction. Until the accreted mass has sufficient counter-angular momentum to destroy the accretion disk, the accretion disk launches jets in the same direction. 

To summarize, I suggest that in some cases of large angular momentum perturbations in the pre-collapse core, which are further amplified by instabilities above the NS (Section \ref{sec:intro} for these processes in the JJEM), there is a positive feedback such that the jets force further accretion to maintain the accretion disk in the same plane, hence the launching of the jets is in the same axis and long-lived. The two opposite jets might form a prominent symmetry axis. Eventually, a large perturbation with angular momentum in the opposite direction to the accretion disk will destroy the disk. The next disk will have a different plane. 
This positive-direction-jet-feedback mechanism is the subject of future three-dimensional hydrodynamical simulations. 

% ==================================
\section{Discussion and Summary} 
\label{sec:Summary}
% ==================================

I use the JJEM prediction that jittering jets shape many CCSNRs to have point-symmetric morphologies to search for one in the complex 320.4-1.2, which contains the nebula MSH 15-52 around the pulsar PSR B1509-58 and the RCW 89 CCSNR. I identified a point-symmetric morphology of two main symmetry axes in the new radio observations of RCW 89 (Figure \ref{Fig:RCW89FigureRadioCen}). The long axis that the inner radio image defines also extends to a bright radio peak in the northeast (Figure \ref{Fig:RCW89FigureRadioExtend}) and along the optical extension to the northeast (Figure \ref{Fig:RCW89Optical}). I use the identification of a point-symmetric morphology in RCW 89 and the success of the JJEM in recent years in explaining point-symmetric CCSNRs to argue that RCW 89 is a separate CCSNR from the pulsar wind nebula MSH 15-52 (Section \ref{subsec:RCW89}); earlier studies (Section \ref{subsec:Early}) considered these two regions to be one CCSNR. 

In Section \ref{subsec:PWN}, I tentatively speculate on a possible scenario where the two CCSNRs occurred within a short time and have been interacting with each other. The interaction involves four main phases. (1) For tens of thousands of years, the progenitor of RCW 89 blew an intense wind around the progenitor of the pulsar. (2)  The progenitor of the pulsar exploded as a CCSN, and the ejecta interacted with the dense wind. (3) Several hundred years later, the progenitor of RCW 89 exploded. The interaction of the ejecta from the two CCSNRs is highly unstable, forming fingers of the X-ray hand-like, `cosmic hand' structure around the pulsar. (4) The pulsar wind nebula filled the hand-like-shaped cavities, explaining the hard X-ray emission of this structure. This speculative scenario deserves a detailed study, including hydrodynamical simulations.

In this study, I utilized the JJEM to investigate a CCSNR, RCW 89, and concluded that it is a separate CCSNR from MSH 15-52. Some recent studies use the alternative neutrino-driven mechanism to explore other properties of remnants. Two recent studies used the neutrino-driven mechanism to study properties of the compact object remnants, namely, the NS or black hole remnants.  \cite{Willcoxetal2025} study the masses of black hole remnants in binary systems, and \cite{Tsunaetal2025} study the fate of post-main-sequence binary mergers. Both studies accept the prediction of the neutrino-driven mechanism concerning `failed supernovae', i.e., that some massive stars failed to explode and leave black hole remnants. In the JJEM, there are no `failed supernovae.' There is an explosion even when a black hole is formed, an explosion that can be very energetic (e.g.,  \citealt{Gilkisetal2016}). Recent observations that find no evidence for `failed supernovae' at the rate the neutrino-driven mechanism predicts (e.g., \citealt{ByrneFraser2022, StrotjohannOfekGalYam2024, BeasoretalLuminosity2025, Healyetal2025}), support the JJEM in this specific dispute.  

There are several other CCSNRs in addition to RCW 89 that present two or three main symmetry axes, which the JJEM attributes to powerful pairs of jets. In Section \ref{sec:PowrfulPair} I proposed a positive-direction-jet-feedback mechanism, where energetic jets that inflate two large bubbles (cavities) force further accretion to have angular momentum along the same axis as the jets. This prolongs the life of the pair of jets. 

This paper contributes to the rich variety of processes and applications of the JJEM, thereby strengthening the JJEM.

% ======================================
\section*{Acknowledgements}
% ======================================

I thank Manuel C. Peitsch for providing panel (b) of Figure \ref{Fig:RCW89Optical}. {{ I thank an anonymous referee for helpful comments that improved the presentation of the results.  }} This research has made use of \textit{Aladin sky atlas} developed at CDS, Strasbourg Observatory, France (\citealt{Aladin1, Aladin2, aladin3}).
I thank the Charles Wolfson Academic Chair at the Technion for the support.

% =================================
% =================================
% =================================

%\bibliographystyle{mnras}
%
%  For ApJ style: 
% \bibliography{reference}{}
% \bibliographystyle{aasjournal}
% =================================
% =================================
%%%% Insert here the references list 

% =================================
% =================================
 
  \end{document}